\documentclass[aps,prl,superscriptaddress,twocolumn,balancelastpage]{revtex4-2}

\usepackage[colorlinks,bookmarks=false,citecolor=blue,linkcolor=blue,urlcolor=blue]{hyperref}
\usepackage[all]{hypcap}   

\usepackage{amsmath,amssymb}
\usepackage{graphicx}
\usepackage{enumerate}

\usepackage{verbatim}
\usepackage{color}

\usepackage{placeins}  

\usepackage{braket}
\usepackage{dsfont}

\usepackage{physics}

\usepackage{enumitem}

\newcommand{\ue}{\text{e}}
\newcommand{\ui}{\text{i}}

\newcommand{\ua}{\text{A}}
\newcommand{\ub}{\text{B}}
\newcommand{\uc}{\text{c}}
\newcommand{\uH}{\text{H}}
\newcommand{\uSH}{\text{SH}}

\newcommand{\U}{\mathcal{U}}

\newcommand{\HIDDEN}[1]{}

\makeatletter
\let\Hy@backout\@gobble
\makeatother

\begin{document}

\title{Universal spectral correlations in interacting chaotic few-body quantum systems}

\author{Felix Fritzsch}
\affiliation{Physics Department, Faculty of Mathematics and Physics,
    University of Ljubljana, Ljubljana, Slovenia}

\author{Maximilian F. I. Kieler}
\affiliation{Technische Universit\"at Dresden,
 Institut f\"ur Theoretische Physik and Center for Dynamics,
 Dresden, Germany}

\date{\today}

\begin{abstract}
    The emergence of random matrix spectral correlations in interacting quantum 
    systems is a defining feature of quantum chaos.    
    We study such correlations in terms of the spectral form factor in interacting chaotic few- and
    many-body systems, modeled by suitable random-matrix ensembles,
    and obtain exact results for large Hilbert space dimensions.
    The transition of the spectral form factor from the non-interacting to the strongly interacting case can be described as a simple combination of these two limiting cases, which we confirm by extensive numerical studies in few-body systems.
    This transition is universally governed by a single scaling parameter.
    Moreover, our approach accurately captures spectral correlations in actual physical system, which we demonstrate for coupled kicked rotors.
\end{abstract}

\maketitle

The quantum chaos conjecture \cite{CasValGua1980,Ber1981b,BohGiaSch1984} predicts statistical properties of energy levels in quantum systems whose classical limit is chaotic to follow random matrix theory \cite{Dys1962b,Wig1967,Meh1991}.
Using semiclassical periodic orbit theory this connection has been shown to follow from only a few basic properties of the chaotic classical dynamics \cite{Gut1990,SieRic2001,Sie2002,MueHeuBraHaaAlt2004}.
Subsequently random-matrix like spectral statistics has become one of the most 
widely used definitions of quantum chaos even in the absence of a classical 
limit.
A distinguished feature of the spectrum of such chaotic quantum 
systems and the corresponding random matrix ensembles is the 
presence of correlations between energy levels in contrast to 
the uncorrelated Poissonian spectrum of integrable 
\cite{BerTab1977} or (many-body \cite{PalHus2010}) localized 
systems \cite{AltShk1986,AltZhaKotShk1988}.
These correlations are conveniently detected by the spectral form factor (SFF) \cite{Haa2010} which has received growing attention in recent years in, e.g., high energy physics \cite{CotHunLiuYos2017,CotGurHanPolSaaSheStaStrTez2017,GhaHanSheTez2018,WinSwi2022} as well as condensed matter and 
many-body systems \cite{KosLjuPro2018,BerKosPro2018,BerKosPro2021,CheLud2018,SunBonProVid2020,ChaDeCha2018b,ChaDeCha2018,FriChaDeCha2019,GarCha2021,ChaDeCha2021,BerKosPro2022,DagMisChaSad2022:p,WinBarBalGalSwi2022a,BarWinBalSwiGal2023:p,FlaBerPro2020,KosBerPro2021,MouPreHusCha2021,AkiWalGutGuh2016}.
Recently, the SFF has been shown to follow random matrix theory in various solvable instances of chaotic many-body systems involving both homogeneous \cite{KosLjuPro2018,BerKosPro2018,BerKosPro2021} and random quantum circuit models \cite{ChaDeCha2018b,ChaDeCha2018,FriChaDeCha2019,GarCha2021,ChaDeCha2021}.
The latter constitute random matrix ensembles which incorporate the (spatial) locality of typical many-body systems. 

We study the SFF in a similar random matrix model, which is built from large independent chaotic subsystems subject to an all-to-all, and hence spatially non-local, interaction of tunable strength.
For the bipartite case of just two subsystems our setting reduces to the so-called random matrix transition ensembles (RMTE) introduced in Ref.~\cite{SriTomLakKetBae2016}.
We therefore refer to our setting as the extended RMTE henceforth.
The bipartite RMTE models a universal transition from an uncorrelated Poissonian spectrum with exponentially distributed level spacings in the non-interacting case to a correlated spectrum whose spacings follow Wigner Dyson statistics at strong interaction \cite{SriTomLakKetBae2016}.
This universal transition has been observed subsequently also in the average eigenstate entanglement \cite{LakSriKetBaeTom2016,TomLakSriBae2018,HerKieFriBae2020} and in the entanglement generation after a quench \cite{PulLakSriBaeTom2020,PulLakSriKieBaeTom2023}.

In the extended RMTE we describe the full transition of the SFF from a simple 
product structure in the non-interacting case towards the full random matrix 
result at strong interaction as a simple convex combination of these two extreme cases.
This transition is universally governed by a single scaling parameter, which combines the dependence of the SFF on all parameters of the system, namely the interaction strength as well as the size and the number of subsystems, into a single number.
The SFF signals an intricate interplay between different time (and associated energy) scales, such as the Heisenberg time of the subsystems and the full system, and most notably a non-trivial Thouless time. 
We confirm our prediction by extensive numerical studies in few-body systems but expect our results to hold even in the many-body case.
For the minimal setting of a bipartite system, we obtain a similar description 
also for the moments of the spectral form factor, which characterize its 
distribution and indicate correlations between multiple levels.
Moreover we go beyond random matrix models and demonstrate that the above results equally well apply in quantized dynamical systems, i.e., a pair of coupled kicked rotors.
Ultimately, we complement our results by a perturbative treatment of the interaction.

\emph{Extended random matrix transition ensemble.}---To model interacting few- or many-body 
systems, we generalize the RMTE
introduced in Ref.~\cite{SriTomLakKetBae2016} by allowing for an arbitrary 
number $L$ of subsystems.
We consider Floquet systems which evolve in discrete time with unitary time evolution operator  given by
\begin{align}
\U = \U_\uc(\epsilon) \left(\U_1 \otimes \U_2 \otimes \cdots \otimes \U_L 
\right).
\label{eq:rmte}
\end{align}
Each of the $\U_i$ is an independent, $N$-dimensional Haar random unitary drawn 
from the circular unitary ensemble, CUE$(N)$, and models a chaotic subsystem.
By considering the CUE we restrict ourselves to systems without anti-unitary symmetries, 
e.g., time-reversal invariance.
The interaction is introduced by the $N^L$-dimensional diagonal unitary matrix 
$\U_\uc(\epsilon)$, where $\epsilon$ controls the strength of the interaction 
with $\epsilon=0$ corresponding to the non-interacting situation, $\U_\uc(0) = 
\mathds{1}$. In the canonical product basis the interaction reads
\begin{align}
\left[\U_\uc(\epsilon)\right]_{j_1 \cdots j_L}^{i_1\cdots i_L} = \delta^{i_1}_{j_1}\cdots\delta^{i_L}_{j_L} \exp\left(\ui \epsilon \xi_{i_1\cdots i_L}\right).
\end{align}
Here the phases $\xi_{i_1\cdots i_L}$ are i.i.d.~random variables with zero mean and variance $\sigma^2$, which gives rise to an effective interaction strength $\sigma \epsilon$.
For numerical simulations we use phases uniformly distributed in $\left[-\pi,\pi\right]$.
Note, that imposing a spatial locality structure on $\U_\uc$ recovers the random-phase circuit of Ref.~\cite{ChaDeCha2018}. 

\emph{Spectral form factor.}---The SFF indicates correlations between the eigenphases (or quasi-energies) $\phi_i$ defined by the eigenvalue equation $\U \ket{i} = \exp\left(\ui \phi_i\right)\ket{i}$.
For chaotic Floquet systems the spectral density is a constant, but its two-point 
correlation function $r_2(\omega)$ yields the probability of finding two 
eigenphases with a distance $\omega$ and hence encodes spectral correlations.
The SFF $K(t)$ is then given by the Fourier transform of the connected part of 
$r_2(\omega)$ and depends on a time variable $t$ conjugate to the quasi-energy 
difference $\omega$.
The SFF has a simple representation in terms of the time evolution operator $\U$ as
\begin{align}
K(t) = \big\langle | \text{tr}\left(\U^t\right) |^2 \big\rangle  - N^{2L}\delta_t^0.
\label{eq:spectral_form_factor_trace}
\end{align}
Here, the brackets denote an ensemble average over the subsystems, i.e., over 
the $L$ independent CUE$(N)$, as well as an average over the random phases 
$\xi_{i_1 \cdots i_L}$.
For numerical simulation we average over at least 1000 realizations.
This averaging procedure is necessary as the SFF is not self averaging 
\cite{Pra1997} and fluctuates wildly for an individual realization.

It is instructive to begin with the SFF for a single CUE of dimension 
$M$, for which the SFF takes the simple form $K_M(t) = \min\{t, M\}$.
The initial linear ramp $\sim t$ indicates correlations in the spectrum and 
substantially differs from the constant SFF of an uncorrelated Poissonian 
spectrum, characteristic for, e.g., integrable systems 
\cite{BerTab1977}.
Hence a linear ramp of the SFF indicates quantum chaos and ergodicity.
In interacting physical models the linear ramp is usually approached after a non-universal time scale known as Thouless time $t_{\text{Th}}$.
It sets the energy scale $\sim 1/t_{\text{Th}}$ below which the system exhibits random matrix like spectral correlations and hence indicates the onset of universal dynamics.
In contrast for non-interacting systems modeled by the tensor product of independent CUE$(N)$ matrices, e.g., the extended RMTE at $\epsilon=0$, the SFF factorizes into a product $K(t) = \left[K_N(t)\right]^L$.

\begin{figure}[]
    \centering
    \includegraphics[width=8.5cm]{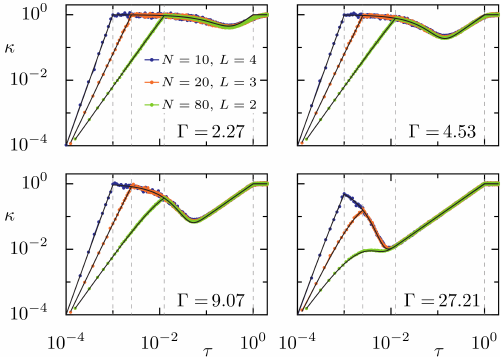}
    \caption{SFF $\kappa(\tau)$ for the extended RMTE for 
        different $N$, $L$ and $\Gamma$. Black lines correspond 
        to 
        Eq.~\eqref{eq:spectral_form_factor_semiclassics2}. 
        Dashed 
        gray lines correspond to $\tau_\uSH$ and $\tau_\uH$.}
    \label{fig:sff_rmte_universality}
\end{figure}

In the extended RMTE we expect a transition from this factorized SFF to the full CUE$\left(N^L\right)$ SFF for increasing interaction strength.
In the following we fully characterize this transition and demonstrate that it depends on a single scaling parameter only.
To this end we adapt the large $N$ expansion of the SFF for the random-phase circuit of Ref.~\cite{ChaDeCha2018} based on the Weingarten calculus for integration over unitary groups \cite{Col2003,ColSni2006} to the extended RMTE.
The average over the subsytems proceeds in the same fashion whereas the average over the phases simplifies; see Ref.~\cite{FriKie2023_long} for a detailed derivation.
Ultimately, as our first main result, we represent the SFF in the simple form of a time-dependent convex combination of the two extreme cases discussed above.
It is given by 
\begin{align}
K(t) = |\chi(\epsilon)|^{2t}K_N(t)^L + \left(1 - |\chi(\epsilon)|^{2t}\right) K_{N^L}(t),
\label{eq:spectral_form_factor_semiclassics2}
\end{align}
where $\chi(\epsilon)=\langle \exp\left(\ui \epsilon \xi\right) \rangle_\xi$ is the characteristic function of the distribution of the phases $\xi_{i_1\cdots i_L}$.
This result is exact in the limit $N \to \infty$.
For finite $N$ it provides the leading contribution (in $1/N$) for times $t<t_{\text{SH}}=N$, i.e., smaller than the subsystems' Heisenberg time $t_{\text{SH}}$ set by the mean level spacing $2\pi/N$ of the subsystems.
For this times $K_N(t)=K_{N^L}(t)=t$.
It has a natural extension to larger times by including the plateaus of $K_N(t)=N$ for $t>t_\uSH$ and $K_{N^L}(t)=N^L$ for times $t>t_\uH=N^L$, i.e., larger than the full systems Heisenberg time $t_\uH$.
This extension is an approximation, which is in excellent agreement with numerical data as depicted in Fig~\ref{fig:sff_rmte_universality} with possible deviations occurring around Heisenberg time and for small coupling.
We emphasize, that requiring large $N$ limits numerical studies to few-body systems, i.e., small $L$, while our arguments do not depend on $L$ being small.
We therefore expect our results to hold also in the many-body setting.
However, before discussing the qualitative features of the SFF 
in the RMTE in more detail, we first point out its universal 
dependence on a single scaling parameter.

\emph{Universality.}---To compare the SFF for different systems it is appropriate to measure both $K(t)$ and time $t$ in units of $t_\text{H}$ and to introduce the rescaled SFF 
$\kappa(\tau)$  and the rescaled time $\tau$ via
\begin{align}
    \kappa(\tau) = K(t)/N^L \quad \text{and } \, \tau = t/N^L.
\end{align}
This results in a rescaled Heisenberg time $\tau_\uH=1$ of the full system and $\tau_{\uSH}=N^{-L+1}$ of the subsystems.
Apart from the latter, the only 
$N$ dependence is implicitly contained in $|\chi(\epsilon)|^{2t}$ via $t = N^L 
\tau$.
By applying the central limit theorem to the characteristic function the $N$ dependence together with the dependence on the effective coupling strength $\sigma \epsilon$ can be converted into the dependence on a single scaling parameter $\Gamma$ via
 \begin{align}
 |\chi(\epsilon)|^{2t} = \exp\left(-\Gamma^2\tau\right) \quad \text{with } \, \Gamma=\sigma \epsilon N^{L/2}.
 \label{eq:central_limit_theorem}
 \end{align}
Here we use the characteristic function $\exp\left(-x^2/2\right)$ of the standard normal 
distribution.
Consequently, the SFF becomes independent from the concrete choice of the distribution of the phases $\xi_{i_1 \cdots i_L}$ entering $\U_\uc$.
Moreover, it depends only on $\Gamma$ for times $\tau > \tau_\uSH$.
This universal dependence on a single scaling parameter constitutes our second main result.
It is well confirmed in Fig.~\ref{fig:sff_rmte_universality}, where we depict the SFF for different combinations of $N$, $L$, and $\epsilon$ all leading to the same $\Gamma$ and coinciding SFF for $\tau > \tau_{\uSH}$.
In the non-interacting case $\Gamma=0$ and hence $\exp\left(-\Gamma^2\tau\right)=1$
the SFF initially grows as $\kappa(\tau)=\tau^{L}$ up to $\tau=\tau_{\uSH}$ and 
subsequently is constant, $\kappa(\tau)=1$ (not shown).
For small $\Gamma$ we still observe an initial growth of the SFF as $\kappa(\tau)\sim \tau^L$, but after times larger than $\tau_\uSH$ the SFF drops down to the linear ramp $\kappa(\tau) \sim \tau$ because all other terms are exponentially suppressed as $\exp\left(-\Gamma^2\tau\right)$.
This indicates the Thouless time $\tau_{\text{Th}}$ as the smallest time for which $\kappa(\tau) \sim \tau$.
For intermediate $\Gamma$ one has $\tau_\uSH<\tau_{\text{Th}}<1$ and we obtain \cite{FriKie2023_long}
\begin{align}
t_{\text{Th}}= N^L\tau_{\text{Th}}= \frac{L\ln(N)}{2|\ln|\chi(\epsilon)||},
\label{eq:Thouless_time}
\end{align}
which scales linear with the number of subsystems.
This is in contrast with, e.g. logarithmic scaling \cite{KosLjuPro2018,ChaDeCha2018,GarCha2021} for local interactions or even $t_\text{Th}=0$ in local dual-unitary quantum circuits \cite{BerKosPro2018, BerKosPro2021}.
For large $\Gamma$ the linear ramp is approached earlier than $\tau_\uSH$, as shown for $\Gamma=27.21$ for $N=80$ and $L=2$.
Ultimately for very large $\Gamma$ all terms involving the 
characteristic function are almost immediately suppressed and 
the SFF reduces to the CUE$\left(N^L\right)$ result (not shown).

\begin{figure}[]
    \centering
    \includegraphics[width=8.5cm]{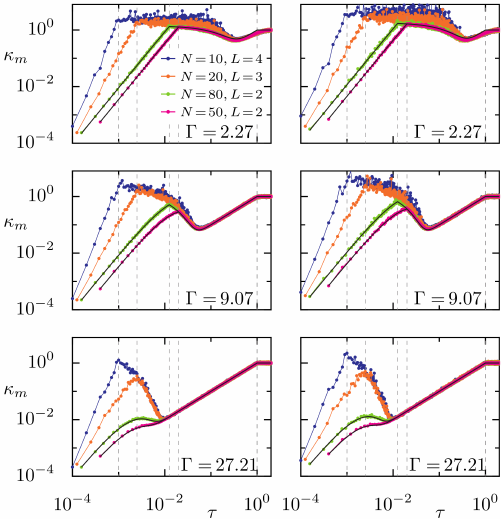}
    \caption{Second ($m=2$ ,left) and third ($m=3$, right) 
        moment of the SFF $\kappa_m(\tau)$ for the extended RMTE 
        for different $N$, $L$ and $\Gamma$. Black lines are obtained 
        from Eq.~\eqref{eq:moments_semiclassics_1}. Dashed gray 
        lines correspond to $\tau=\tau_\uSH$ and $\tau=\tau_\uH$.}
    \label{fig:sff_moments23_rmte}
\end{figure}

\emph{Higher moments.}---As the SFF is defined via an average over the RMTE one might study its distribution via its moments of order $m$ defined by
\begin{align}
K_m(t) = \big\langle \big| \text{tr}\left(\U^t\right) \big |^{2m} \big\rangle  - N^{2Lm}\delta_{t}^0.
\label{eq:moments_sff}
\end{align}
For the CUE$(M)$ the SFF follows an exponential distribution, i.e., 
$K_{M,m}(t) = m!K_ {M}(t)^m$ \cite{Kun1999}.
To compute the moments in the extended RMTE for $t<t_\uSH$ we follow Ref.~\cite{ChaDeCha2021} to perform the 
average over the independent CUE$(N)$.
The remaining average over the phases $\xi_{i_1\cdots i_L}$ yields \cite{FriKie2023_long}
\begin{align}
K_m(t)  = m!t^m\sum_{k=0}^{m}A_k(t)|\chi(\epsilon)|^{2t(m - k)}
\label{eq:moments_semiclassics_1}
\end{align}
for initial times $t<t_\uSH$.
Here the combinatorical factors $A_k(t)$ are polynomials of degree $m(L-1)$ in $t$ which can be obtained exactly only for the bipartite case $L=2$.
Computing the latter for $L=2$ and fixed $m$ allows for expressing the SFF as a time dependent convex combination between the full random matrix result $K_{N^2,m}(t)$ and the non-interacting result $\left[K_{N,m}(t)\right]^2$ as well as additional terms involving products of lower moments.
For instance for the second moment, $m=2$, we find \cite{FriKie2023_long}
\begin{align}
K_2(t) =& \,  K_{N^2,2}(t) \left(1 - |\chi(\epsilon)|\right)^2 + \left[K_{N,2}(t)\right]^2 |\chi(\epsilon)|^{4t} \nonumber \\
& + 4K_{N^2}(t)K_N(t)^2|\chi(\epsilon)|^{2t}\left(1 - |\chi(\epsilon)|^{2t}\right)
\label{eq:moments_semiclassics_3}
\end{align}
and similar for $m>2$.
By explicitly including the plateaus for the moments of the CUE spectral form factors the above results again extends also to times $t>N$.
Moreover, it reproduces the correct result for the non-interacting case $\epsilon=0$ for all $m$ and for the interacting case implies $K(t) \sim K_{N^2,m}(t)$, i.e., an exponential distribution, for $t > t_{\text{Th}}$ as all the terms involving $|\chi(\epsilon)|^{2t}$  have decayed.
Given this exponential distribution we define the rescaled moments via
\begin{align} 
    \kappa_m(\tau) = \frac{1}{N^L}\left(\frac{K_m(t)}{m!} \right)^{1/m}.
    \label{eq:moments_sff_rescaled}
\end{align}
Repeating the argument invoking the central limit theorem, we again find that the  rescaled moments of the SFF depend only on $\Gamma$ for times $\tau > \tau_\uSH$.
Both Eq.~\eqref{eq:moments_semiclassics_3} and its variants for $m>2$, see \cite{FriKie2023_long}, as well as the universal dependence on $\Gamma$ for fixed $L=2$ is confirmed in Fig.~\ref{fig:sff_moments23_rmte} for the second and third moment.
Due to the rescaling~\eqref{eq:moments_sff_rescaled} higher moments exhibit the same phenomenology as the SFF $\kappa(\tau)$.
They depend on both $L$ and $\Gamma$ even for times $\tau_\uSH < \tau \lesssim \tau_{\text{Th}}$ while coinciding with the CUE$(N^L)$ result afterwards.

\begin{figure}[]
    \centering
    \includegraphics[width=8.5cm]{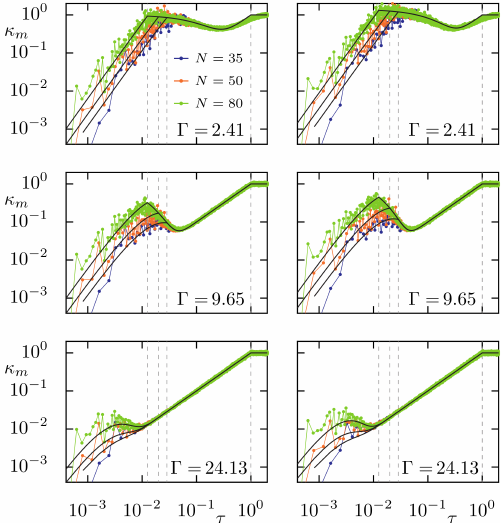}
    \caption{SFF ($m=1$, left) and its second moment ($m=2$, right) $\kappa_m(\tau)$ for the coupled kick rotors for different $N$ and $\Gamma$. Black lines depict the RTME results. Dashed gray lines correspond to $\tau=\tau_\uSH$ and $\tau=\tau_\uH$.}
    \label{fig:sff_kicked_rotor_universality}
\end{figure}

\emph{Coupled kicked rotors.}---To demonstrate, that the RMTE describes actual physical systems, we apply our results to a quantized dynamical system given by two coupled kicked rotors \cite{Fro1972}.
While individual kicked rotors \cite{Chi1979} are a paradigmatic model for both classical and 
single particle quantum chaos, coupling two rotors provides an example for the 
corresponding two-body setting \cite{AdaTodIke1988,Lak2001,GadReeKriSch2013,RicLanBaeKet2014,SriTomLakKetBae2016,
    LakSriKetBaeTom2016,TomLakSriBae2018}.
We consider coupled kicked rotors with periodic boundary conditions, whose 
classical phase space is the four torus with canonical conjugate coordinates 
$(q_1,q_2,p_1,p_2)$.
After quantization the effective Planck's constant $h$ is constraint to 
integer values $1/h = N$.
The time evolution operator is a 
$N^2$-dimensional unitary of the form~\eqref{eq:rmte} with \cite{BerBalTabVor1979,HanBer1980,ChaShi1986,KeaMezRob1999,DegGra2003}
\begin{align}
\U_i = \ue^{-\pi \ui N p_i^2} \ue^{-\frac{\ui \gamma 
N}{2\pi}\cos(2\pi q_i )}.
\end{align}
Here $k_1=9.7$ and $k_2=10.5$ governs the strength of the kicks end ensures chaotic classical dynamics.
The coupling is introduced by
\begin{align}
\U_\uc = 
\ue^{-\frac{\ui \gamma N}{2\pi}\cos(2\pi[q_1 + q_2])},
\end{align}
with coupling strength $\gamma$ and effective $\epsilon=\gamma N /(2\pi)$.
We choose boundary conditions for the quantum states which break time-reversal invariance and average over such boundary conditions in order to perform the average in the definition of the SFF and its moments.
The resulting SFF and its second moment is depicted in Fig.~\ref{fig:sff_kicked_rotor_universality} and shows qualitatively similar behavior as in the RMTE.
However, initial fluctuations are more pronounced, which we attribute to short 
periodic orbits in the classical dynamics.
In order to model the coupled kicked rotors with the bipartite RMTE we choose $\xi_{ij}=\cos(\eta_{ij})$ with 
i.i.d.~and uniformly distributed $\eta_{ij}$.
This yields $\chi(\epsilon) = J_0(\gamma N /(2\pi))$.
The corresponding RMTE result is in good agreement with numerical data and again implies universal dependence on $\Gamma$ for $\tau>\tau_{\uSH}$; see Fig.~\ref{fig:sff_kicked_rotor_universality}.

For scaling parameters $\Gamma$ for which the Thouless time is given by Eq.~\eqref{eq:Thouless_time} we note, that $t_{\text{Th}}$ does not coincide with the Ehrenfest time $t_{\text{E}}$.
The latter is the time it takes for an initially localized wave packet to spread over the system and hence indicates the time for which quantum follows classical dynamics.
It is determined by the classical system's Lyuapunov exponents and for the coupled kicked rotors
approximately reads $t_{\text{E}} \approx \ln(N)/(2\ln(k_\ua 
k_\ub/4))$ \cite{Chi1979}.
For chaotic subsystems $t_{\text{E}}$ is necessarily smaller than the subsystem's Heisenberg time $t=N$ and is also much smaller than $t_{\text{Th}}$ even though both times scale logarithmic with $N$.

\begin{figure}[]
    \centering
    \includegraphics[width=8.5cm]{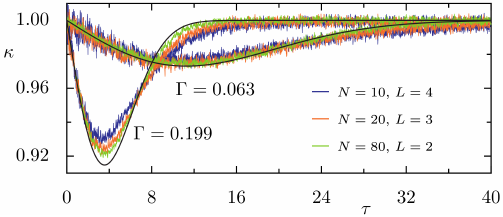}
    \caption{SFF $\kappa(\tau)$ for small $\Gamma$ for 
    $\tau>\tau_{\uSH}$ and different $N,L$.
        We smooth the SFF by an additional moving time average.
     The perturbative result~\eqref{eq:sff_rmte_perturbation_theory} is 
depicted as black lines. Deviations from universality at $\Gamma=0.199$ are of subleading order $1/N$. \label{fig:sff_rmte_perturbation_theory}}
\end{figure}

\emph{Perturbative regime.}---For very small scaling parameter extrapolating the exact result from $t<N$ to larger times gives a less accurate description of the SFF.
This is visible already for $\Gamma = 2.27$ in Fig.~\ref{fig:sff_rmte_universality} around Heisenberg time $\tau \approx \tau_\uH$.
A natural approach for $\Gamma \ll 1$ is to extend the regularized Raleigh-Schr\"odinger perturbation theory introduced in Ref.~\cite{SriTomLakKetBae2016} from the bipartite to the extended case of arbitrary $L$.
Viewing $\U_\uc(\epsilon)$ as a perturbation to the non-interacting system the eigenphases $\phi_i=\phi_i(\epsilon)$ can be expanded in a perturbative series in $\epsilon$ which allows for computing $K(t) = \langle \exp\left(\ui \sum_{ij} \phi_i - \phi_j \right) \rangle$.
While Eq.~\eqref{eq:moments_semiclassics_1} still holds for $\tau < \tau_\uSH$ 
the perturbative approach yields \cite{FriKie2023_long}
\begin{align} 
\kappa(\tau) = 1 - \Gamma^2 \tau \ue^{-(\Gamma \tau)^2}
\label{eq:sff_rmte_perturbation_theory}
\end{align}
for $\tau > \tau_{\uSH}$ up to arbitrary large times.
Again, this universally depends on the scaling parameter $\Gamma$ only.
The validity of the perturbative approach for very small $\Gamma$ is depicted in Fig.~\ref{fig:sff_rmte_perturbation_theory}.

\emph{Summary and Outlook.}---We have given a simple description of the SFF (and its moments for the bipartite case) for interacting chaotic subsystems as a convex combination of the results for the non-interacting and the strongly interacting case.
We confirm this numerically for few-body systems and expect it to hold also for many-body systems at large $N$.
Interestingly relatively small subsystem sizes, $N \approx 10$, seem to be large enough 
for our description to apply.
Our description additionally implies the universal dependence of the SFF on a single scaling parameter $\Gamma$ and is insensitive to the detailed statistics of the phases $\xi_{i_1\cdots i_L}$.
However, using i.i.d.~phases we ignore all correlations in the phases as they would be present for instance due to spatial locality of typical many-body systems.
It therefore is an interesting open question, whether such a simple picture applies also for these situations.
Moreover, our results for the RMTE are exact only for small times $t<N$ whereas a derivation for larger times might be possible using field theoretical methods \cite{Zir1996,AltGnuHaaMic2015}.
For systems originating from the quantization of classically chaotic systems, e.g., the coupled kicked rotors, semiclassical periodic-orbit based techniques might shed further light on spectral correlations.
The latter approaches, however, are left for future research.

\emph{Acknowledgements.}---
We thank A.~B\"{a}cker for insightful discussions.
FF further acknowledges fruitful discussion with P.~Kos, F.~G.~Montoya and T.~Prosen.
The work has been supported by Deutsche Forschungsgemeinschaft (DFG), Project 
No. 453812159 (FF) and Project No. 497038782 (MK).

\appendix

\bibliography{paper_sff_rmte}

\end{document}